\documentclass[%
 aip,
 amsmath,amssymb,
preprint,%
]{revtex4-1}

\usepackage{graphicx}
\usepackage{dcolumn}
\usepackage{bm}

\usepackage[utf8]{inputenc}
\usepackage[T1]{fontenc}
\usepackage{mathptmx}
\usepackage{etoolbox}

\usepackage[T1]{fontenc}
\usepackage[utf8]{inputenc}
\usepackage[absolute,overlay]{textpos}
\usepackage{hyperref}
\usepackage{graphicx}
\usepackage{dcolumn}
\usepackage{bm}
\usepackage{xcolor}
\usepackage{siunitx}
\usepackage[all]{nowidow}

\newcommand*\diff{\mathop{}\!\mathrm{d}}

\newcommand{\blu}[1]{{\color{black}{#1}}}

\makeatletter
\def\@email#1#2{%
 \endgroup
 \patchcmd{\titleblock@produce}
  {\frontmatter@RRAPformat}
  {\frontmatter@RRAPformat{\produce@RRAP{*#1\href{mailto:#2}{#2}}}\frontmatter@RRAPformat}
  {}{}
}%
\makeatother

\begin{document}

\preprint{AIP/123-QED}

\title[]{A Lagrangian Particle-Based Numerical Model for Surfactant-Laden Droplets at Macroscales}
\author{Mateusz Denys}
\affiliation{Institute of Physics, Polish Academy of Sciences, Al. Lotnik\'ow 32/46, 02-668 Warsaw, Poland}
 \email{mateusz.denys@gmail.com}
\author{Piotr Deuar}%
\affiliation{Institute of Physics, Polish Academy of Sciences, Al. Lotnik\'ow 32/46, 02-668 Warsaw, Poland}

\author{Zhizhao Che}
\affiliation{%
State Key Laboratory of Engines, Tianjin University, 300350 Tianjin, China
}%

\author{Panagiotis E. Theodorakis}%
 \email{panos@ifpan.edu.pl}
\affiliation{Institute of Physics, Polish Academy of Sciences, Al. Lotnik\'ow 32/46, 02-668 Warsaw, Poland}

\date{\today}

\begin{abstract}
Atmospheric aerosols can consist of inorganic and organic
substances, including surfactants at a significant
concentration. Importantly, the latter can reduce the surface
tension at the liquid--vapor surfaces, where they
preferentially adsorb due to their amphiphilic structure.
As a result, processes such as droplet coalescence,
development of precipitation and ultimately cloud lifetime,
may depend on the presence of surfactants in the aerosols.
Here, we present a numerical model for cloud droplet
formation, which is based on the Lagrangian particle-based
microphysics-scheme super-droplet method and takes into
account the presence of surfactant in the droplets.
Our results show that surfactant facilitates cloud formation
by increasing the number and size of activated droplets,
which concentrate at the bottom of the cloud, while the
largest droplets are concentrated at the top of the cloud.
This indicates a circulation of droplets that involves
activation and growth processes from the bottom of the cloud
towards the top. Moreover, our conclusions are independent 
of the particular approach used for modeling the diffusion
of Eulerian variables due to the subgrid-scale turbulence.
We anticipate that our results will enrich our understanding
of the role of surfactants in the behavior of atmospheric
aerosols and, importantly, will pave the way for further
developments in the numerical modeling of systems with
surfactants at macroscopic scales. 
\end{abstract}

\maketitle

\section{Introduction}
\label{intro}

Pollutants and other atmospheric aerosols play a crucial 
role in Earth's climate change \cite{charlson1992climate, smith2014two} 
with their study dating back to the 1970s \cite{twomey1974pollution, twomey1977influence}. 
A range of different pollutants have been found to
significantly affect cloud development \cite{koren2004measurement, kaufman2005effect, zhang2011impact},
which lead to changes in their properties,
such as radiative properties \cite{garrett2006increased}.
In particular, humic-like substances (HULIS) play a
significant role in decreasing the surface tension of aqueous
solutions in the atmosphere and influencing cloud-droplet
activation properties \cite{kokkola2006cloud}. 
How exactly these effects emerge in these processes still
remains a topic of intensive research, which, among others,
requires the development of suitable models that can in
detail capture the behavior of such systems, thus providing
a deeper understanding of the fundamental underlying
processes. Not only will such an understanding allow us to
implement better policies that aim at preventing extreme
phenomena as a result of climate changes, such as droughts
and floods, but it will also enable novel technological
developments that require the numerical modeling of additives
in liquid phases, which are relevant in applications,
such as atomization, cooling electronic devices, and
electricity production (\textit{e.g.} steam turbines).

Atmospheric aerosols can consist of various inorganic and
organic substances, with a  part of them being salts and
\emph{surfactants}. The latter are amphiphilic molecules
consisting of a hydrophilic and a hydrophobic part in their
molecular structure \cite{Noziere16,Gerard19,Noziere17,Gerard16,Noziere14,Baduel12,Ekstrom10}. 
The source of surfactants in the systems can be diverse, 
for example, biological \cite{renard2019cloud} as well as
anthropogenic \cite{latif2004surfactants}. Moreover, due to
their amphiphilic character, surfactants preferentially
adsorb at the surface of liquids thus lowering their surface
tension \cite{facchini1999cloud, facchini2000surface, sorjamaa2004role, decesari2005water}, 
and in turn affecting key processes, such as droplet
coalescence, their populations, as well as the development 
of precipitation and ultimately cloud lifetime \cite{sareen2013surfactants}. 
A more thorough understanding of these processes requires
more quantitative descriptions for droplet nucleation and
coalescence phenomena, which are involved in these processes \cite{facchini1999cloud}.
This is crucial as surfactants may affect the activity of cloud condensation nuclei (CCN) \cite{forestieri2018establishing},
\textit{i.e.}, small particles on which water vapor
condensates when the supersaturation of air exceeds a
critical value. In this case, surfactants can contribute to
the mass of a CCN and influence the ability of the aerosol
particles to be activated and form clouds \cite{li1998influence}. 
Further studies have shown that the composition of the
atmospheric CCN significantly affects the cloud droplet
activation, for example, by changing the activation diameter
\cite{raymond2003formation}. 
This takes place by (i) reducing the amount of solute
available in the CCN and (ii) decreasing their surface tension \cite{westervelt2012effect}.
Both effects are expected to play an important role towards
a quantitative description of the cloud formation in the
presence of surfactants, and they, therefore, should be
included in the numerical model.
\blu{Thus, numerical models might provide an indication 
on how the amount and probability of generating rain 
droplets might be affected by the presence/increase of surfactant.
Since much surfactant comes from anthropogenic activities,
numerical models can provide a clue towards understanding
whether the presence of surfactant would favor 
more or less rain. Models also provide information
on which of the properties are expected to be affected
by the presence of surfactant, which eventually affects the
properties of clouds.}

While empirical measurements concerning surfactants in the
atmosphere could ideally provide useful insights into the
process of cloud formation,  numerical studies have emerged as a useful
tool to analyze and understand this process \cite{shima2009super, grabowski2019modeling}. 
For example, such investigations have previously been used to
estimate the influence of surfactants on cloud microphysics
and clouds formation \cite{brimblecombe2004rediscovering, prisle2012surfactant}. However, there is currently no
available simulations of a cloud as a multitude of
surfactant-laden droplets, which could allow for a more
detailed description of the relevant mechanisms.
Although surfactants are a significant part of the
atmospheric aerosol mass, surface tension of aerosol
particles and water droplets has remained unexplored in the
numerical models of the cloud evolution. In fact, most models
traditionally assume a surface tension equal to that of pure
water \cite{bzdek2020surface}. Hence, a numerical model that
incorporates the effect of surface tension might potentially
provide a better understanding of the role of surfactants in
the atmosphere and its influence on cloud evolution. 
Finally, simulating such systems with surfactant at large
(macroscopic) scales with molecular or mesoscopic models
\cite{Theodorakis2015a} would be computationally prohibitive.

The goal of our study is to propose a numerical model that
incorporates the surfactant effect in the coalescence of
droplets at macroscopic scales towards describing phenomena,
such as cloud formation and precipitation. 
Current literature in this area mainly consists of
microscopic and global models with the numerical modeling of
clouds in the presence of surfactants being at its infancy
(see, for example, Sec.\ \ref{lit}). Our numerical model is
described in Sec.\ \ref{assum} with our simulations
suggesting that surfactant may increase the cloud droplet
activation (Sec.\ \ref{res}). Moreover, the model offers
opportunities for further developments of numerical
approaches in this area of cloud physics, which aim at
incorporating the detail of relevant physical phenomena, 
as will be further discussed in Sec.\ \ref{concl}.

\section{Model and methodology}
\label{assum}

\subsection{Background}
\label{lit}

About 20\% of soluble matter in the atmosphere
(even up to 70\% in some cases) consists of organic 
compounds \cite{facchini2000surface}.
Some of these compounds are surface active and can affect
the surface tension of cloud droplets as has been also
highlighted by Bzdek \textit{et al.} \cite{bzdek2020surface}
while investigating the coalescence of picoliter droplets.
Although such a reduction may depend on the droplet 
size \cite{bzdek2020surface}, this still remains under debate
in the context of cloud droplets \cite{morris2015humidity}.
A study by Facchini \textit{et al.} \cite{facchini1999cloud},
which is based on measurements on vacuum-evaporated samples
of cloud water from the Po Valley in Italy,
suggests a large decrease in surface tension with the
concentration of organic solutes that is expected to persist
in growing droplets, as a result of droplet condensation. 
In particular, such large surface-tension changes are crucial
in the case of cloud droplets near the critical size of
nucleation, since they can lead to an increased droplet
population and hence cloud albedo.

The process of cloud-droplet condensation may be described
by K\"ohler's theory \cite{kohlertheory}, which includes two
basic effects: The Kelvin effect, which expresses the change
in the saturation vapor pressure on a curved surface, 
and Raoult's law, which expresses the change in the
saturation vapor pressure in the mixture. More specifically,
the K\"ohler equation for the equilibrium water vapor
saturation ratio, $S$, \textit{vs.} radius, $r$, of a
spherical solution droplet, in Ostwald--Freundlich form is
as follows \cite{prisle2010surfactants}:
\begin{equation}
S \equiv \frac{p_w}{p^0_w} = a_w \exp{\left( \frac{2 \nu_w \sigma}{R T r} \right)},
\label{koehler}
\end{equation}
where $p_w$ is the equilibrium partial pressure of water over
the solution droplet (actual vapor pressure), $p^0_w$
is the saturation vapor pressure over a flat surface of pure
water, $a_w$ is the droplet solution water activity, 
$\nu_w$ is the partial molar volume of water in the solution,
$\sigma$ is the droplet surface tension, 
$R$ is the universal gas constant, 
and $T$ is the Kelvin temperature. The exponential term of
Eq.\ (\ref{koehler}) is the Kelvin term. The water activity
\begin{equation}
a_w = \frac{r^3 - r_d^3}{r^3 - r_d^3 (1 - \kappa)}
\end{equation}
in Eq.\ (\ref{koehler}) is Raoult's term, 
where $r_d$ is the radius of the dry part of a droplet, 
and $\kappa$ is the hygroscopicity of the solute.
Raoult's Law states that $p_w = a_w p^0_w$. 

Ovadnevaite \textit{et al.} \cite{ovadnevaite2017surface}
have theoretically and empirically proven that the reduction
of surface tension can prevail over the reduction in the
Raoult effect in the case of ambient air, which leads to a
substantial increase in the cloud droplet concentration.
Furthermore, Sorjamaa \textit{et al.} \cite{sorjamaa2004role}
have shown that the presence of salt in a droplet drives
surfactant to the surface, which reduces Raoult's term,
while leaving the Kelvin term still considerable. 
At the same time, Prisle \textit{et al.} \cite{prisle2008surfactant, prisle2010surfactants, prisle2012surfactant}, 
by using a global circulation model, have found that if 
K\"ohler theory does not account for surfactant partitioning
and only makes use of the reduced surface tension,
it underpredicts the empirical critical supersaturation.
Moreover, according to Ref.~\cite{petters2013single}, 
the decrease of the hygroscopicity, due to the presence of
surfactant, can partially or fully compensate for the
surface tension reduction occurring at the droplet interface.
Therefore, both the Kelvin effect and Raoult's law can be
considered in a numerical model of a cloud. 
However, before taking such a step, here, we investigate 
the role of surface tension (Kelvin term), and leave a full
investigation of Raoult's term for our future work. 

Lo \textit{et al.} \cite{lo1996effect} have investigated the
influence of surfactants on the droplet evolution. 
They have used sodium dodecyl sulfate (SDS), 
a surrogate of the natural surfactants, and examined its
influence on artificial fog droplets in hydrophobic organic
compounds (HOCs) enrichment reactor. 
There are also numerous simulation models of a single cloud
formation and evolution, for example, 
see Ref.~\cite{shima2009super} and references therein.
However, none of the above studies have taken into account
the presence of surfactants, although these are expected 
to influence the underlying complex mechanisms \cite{mcgraw2021surfactants}. 
To the best of our knowledge, our endeavor is the first
attempt to address this issue by building a
simulation model that includes the effect of surfactant in
the cloud evolution, such as the formation of clouds and the
amount of precipitation.

\subsection{Super-droplet method}

Different cloud dynamics models have been thus far proposed
in the literature, such as the bulk parametrization method
\cite{kessler1969distribution} or the spectral (bin) method
\cite{clark1973numerical}
(see also Ref.~\cite{khain2015representation} and references
therein). These models are based on the Eulerian approach,
where the cloud properties are continuous in space without
considering single cloud droplets. Here, a rather recent
approach to accurately model the cloud dynamics will be used,
that is the Lagrangian particle-based microphysics scheme
Super-Droplet Method (SDM) \cite{shima2009super}. 
Our study extends the capabilities of this model by 
including the numerical modeling of surfactant effects in 
the SDM model, thus providing a possible direction for
future developments in the area of numerical modeling
of systems with droplets at macroscopic scales.

In the Lagrangian approach, computational particles are cloud
droplets \cite{arabas2015libcloudph++}, which connect the
large-scale simulation with the underlying microscopic fluid
mechanics. A main advantage of this approach is avoiding the
need of representing each individual droplet of the system
in the computational model, which would translate into a huge
computational cost. Instead, in the SDM, aggregates of single
droplets are considered, which are known as
\emph{super-droplets}. Hence, the super-droplet can be
treated as an ensemble of many ordinary droplets with the
same intrinsic parameters. A multiplicity parameter indicates
the number of those single droplets represented by the
super-droplet, which is denoted by the positive integer
$\xi_i(t)$. The multiplicity parameter can be different in
each super-droplet and time-dependent, for example, due to
droplet coalescence events. Moreover, each super-droplet has
its own position \textbf{x}$_i(t)$ and attributes
\textbf{a}$_i(t)$, which characterize the $\xi_i(t)$
identical droplets represented by the super-droplet $i$.
For instance, attributes of the super-droplet can include the
radius and the solute mass, \textit{i.e.}, 
\textbf{a}$_i(t) = [R_i(t), M_i(t)]$ \cite{shima2009super}.
In other words, a super-droplet reflects a coarse-grained
representation of single droplets both in real and
attribute space. 

SDM involves a coalescence scheme of super-droplets,
which is outlined as follows: 
\begin{figure}
    \centering
    \includegraphics[width=0.7\columnwidth]{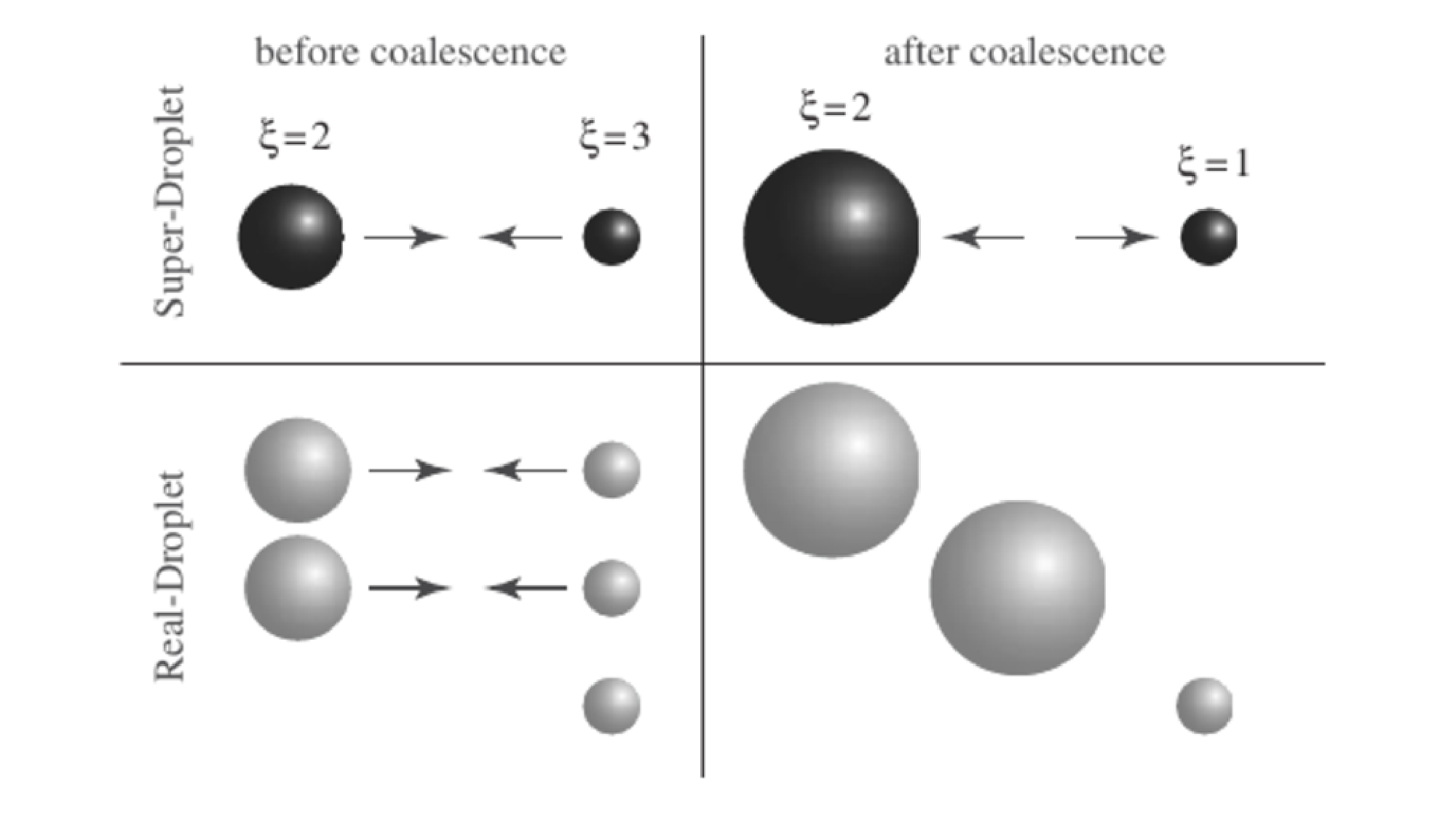}
    \caption{Scheme of coalescence for SDM.
    \cite{shima2009super} Top: coalescence of two
    super-droplets with initial multiplicity 2 and 3. 
    Bottom: the corresponding scheme for the coalescence of
    real droplets. Left: before coalescence.
    Right: after coalescence
    (multiplicities are now 2 and 1).
    Copyright © 2009 by Royal Meteorological Society. All rights reserved.}
    \label{sdm}
\end{figure}
\begin{enumerate}
\item If $\xi_j \neq \xi_k$, we can choose $\xi_j > \xi_k$
without losing generality (Fig.~\ref{sdm}).
Then, the prime variables represent the updated values 
after coalescence.
\begin{eqnarray}
\xi'_j = \xi_j - \xi_k, & \quad & \xi'_k = \xi_k, \\
R'_j = R_j, & \quad & R'_k = (R_j^3 + R_k^3)^{1/3}, \\
M'_j = M_j, & \quad & M'_k = (M_j + M_k), \\
x'_j = x_j, & \quad & x'_k = x_k.
\end{eqnarray}
\item If $\xi_j = \xi_k$,
\begin{eqnarray}
& \xi'_j = [\xi_j/2], \quad \xi'_k = \xi_j - [\xi_j/2], \\
& R'_j = R'_k = (R_j^3 + R_k^3)^{1/3}, \\
& M'_j = M'_k = (M_j + M_k), \\
& x'_j = x_j, \quad x'_k = x_k,
\end{eqnarray}
where $[\cdot]$ is the greatest integer less than or
equal to the quantity in the brackets.
\end{enumerate}

The number of super-droplets decreases due to coalescence
only when $\xi_j = \xi_k = 1$, \textit{i.e.}, 
both super-droplets are real droplets. 
Then $\xi_j = 0$ and $\xi_k = 1$, and the super-droplet, $j$,
is removed from the system. Therefore, the number of
super-droplets (which corresponds to the accuracy of SDM)
slowly decreases during the simulation towards an approximate
number of super-droplets that provides the flexible response
of the model to big changes in the number of real droplets.
We define the probability that super-droplets $j$ and $k$
that are present inside a small region $\Delta V$ will
coalesce in a short time interval $(t, t + \Delta t)$ as:
\begin{equation}
P^{(s)}_{jk} = \max(\xi_j, \xi_k) P_{jk}.
\end{equation}
where $P_{jk}$ is the probability that the (real, single)
droplets $j$ and $k$ inside a small region $\Delta V$ will
coalesce in a short time interval $(t, t + \Delta t)$ 
(see Eq.\ (3) in Ref.\ \cite{shima2009super} for the
definition of $P_{jk}$ in the SDM). Such a definition of a
coalescence probability ensures that the expectation value 
of the number of coalescences of real droplets is conserved
when we use the SDM. Moreover, for a coalescence process we
use a reduced set of non-overlapping pairs of super-droplets,
instead of all possible pairs, which further reduces the
computational cost of a simulation \cite{shima2009super}.

Except for the coalescence process, each super-droplet
behaves in just the same way as a droplet. 
Thus, the condensation/evaporation process of a super-droplet
is governed by an appropriate growth equation 
[specifically, Eq.\ (2) in Ref.\ \cite{shima2009super},
cf.\ Eq.\ (\ref{koehler})] and it can 
move according to the equation of motion:
\begin{equation}
\frac{\diff \mathbf{r}_i}{\diff t} = \mathbf{U}(\mathbf{r}_i) - \hat{\mathbf{z}} v_{\infty}(R_i),
\label{motion}
\end{equation}
where $\mathbf{r}_i$ is the droplet position,
$\mathbf{U}(\mathbf{r}_i)$ is the wind velocity,
$v_{\infty}(R_i)$ stands for the terminal velocity\cite{beard1976terminal},
and $\hat{\mathbf{z}}$ is a unit vector in the vertical direction.

Using such a coarse-grained approach makes the simulation of
a cloud feasible, without much compromise on the accuracy of
the results. Moreover, thanks to the super-droplet method,
introducing the changes of surface tension and their effect
on the droplet properties becomes technically more tractable.
Also, there are studies supporting that SDM might be
considered a better option than the bin method for
investigating nonlinear phenomena, such as turbulence in
clouds \cite{grabowski2021cloud}. Finally, the method might
also be generalized to consider ice particles present in the
case of a storm-cloud \cite{shima2020predicting}.

Still, in view of the greater detail that SDM provides in 
our numerical model, this approach might somewhat require
more computational resources in comparison with other 
methods \cite{shima2020predicting}. Moreover, there is an
associated computational cost for adding new attributes to 
a super-droplet as more super-droplets are needed to cover
the increased dimensionality of the attribute space without
comprising the accuracy of the 
simulation \cite{grabowski2019modeling}.
A way of addressing this issue has been proposed
in Ref.~\cite{grabowski2018lagrangian},
which suggests a simpler and computationally more efficient
approach that is based on the Twomey activation 
scheme \cite{twomey1959nuclei}. In this case, super-droplets
form when a CCN is activated, while they do not exist 
outside a cloud.

\subsection{Surfactant modeling}
\label{surf}

Our SDM calculations took place by appropriate modifications
of the libmpdata++, libcloudph++, and UWLCM
libraries\footnote{The libraries are available under these
links: \url{https://github.com/igfuw/libmpdataxx},
\url{https://github.com/igfuw/libcloudphxx},
\url{https://github.com/igfuw/UWLCM}. 
For an alternative, see \url{https://github.com/darothen/superdroplet}.}. 
We have implemented the following changes to the original
approach for cloud simulation \cite{jaruga2015libmpdata++, arabas2015libcloudph++, dziekan2019university} 
in order to incorporate the effects of surfactant:
(i) Particles (droplets) contain a particular amount of
surfactant and salt, while keeping the ratio of surfactant
and salt always the same for each particle. This ratio is
a model parameter. 
(ii) We have derived the formula for the surface tension 
of a droplet by taking into account the surfactant and salt
partitioning, where surfactant covers the surface of a
droplet until it saturates, 
as in Ref.\ \cite{pan2016controlling}. 
(iii) The modified surface tension is then used for
calculating the heat and vapor diffusion equation,
which contains the Kelvin term. 

The UWLCM library offers two methods for modeling the
diffusion of Eulerian variables due to the subgrid-scale
(SGS) turbulence \cite{dziekan2019university}: 
(i) an implicit large-eddy simulation (ILES) method,
with no explicit parametrization of SGS mixing, but instead,
with numerical diffusion of the advection scheme used to
mimic the SGS turbulence;
(ii) a Smagorinsky scheme \cite{smagorinsky1963general}.
Hence, we performed the analysis for three varied models
of the SGS turbulence to ensure that our conclusions are not model-dependent. 
These models are: the ILES model, the Smagorinsky model,
and the Smagorinsky model with turbulent SGS motion of 
the super-droplets (obtained by adding a random velocity
component $u'_{SD}$, specific to each super-droplet 
[cf.\ Eq.~(\ref{motion}), where every component of the
velocity perturbation evolves as in Eq.\ (10) of 
Ref.\ \cite{abade2018broadening}]. 
In general, introducing SGS turbulence models should 
improve the grid convergence
characteristics \cite{shima2020predicting}. 
We also expected that precipitation rate is going to 
decrease after implementing the influence of the SGS
turbulence on the super-droplets \cite{sato2017grid}.
In the code of the simulation of our model, the microphysics of droplets, which was modified for accounting for the presence of surfactant in each droplet, as given later in this section, is independent of the particular approach for the diffusion of Eulerian variables. Therefore, we were able to apply all the three approaches mentioned above in the same manner, without a necessity of any special implementation. For more details about the implementation of the three approaches, see Ref.\ \cite{dziekan2019university}.

In our model, we have assumed two types of solute in each
droplet: ammonium sulfate [(NH$_4$)$_2$SO$_4$] as
salt \cite{dziekan2019university}, and sodium dodecyl 
sulfate (SDS) as surfactant. 
Moreover, an increase in size of a droplet due to the
absorption of water (condensation) reflects a decreasing
surfactant concentration \cite{facchini1999cloud}.
For this reason, we have assumed that a certain percentage
of the dry part of the droplet is SDS and the rest is
ammonium sulfate \cite{li1998influence}. 
We have also assumed that surfactant can replace the 
salt in CCN. Hence, the dry radii of the super-droplets
statistically remain the same
(cf.\ \blu{Westervelt \textit{et al.}} Ref.\ \cite{westervelt2012effect}). 
Further details regarding the concentration of surfactant
in the droplets will be discussed below. 

The variations of surface tension, $\sigma$, can be
calculated by means of the Szyszkowski equation of 
state \cite{gast1997physical}. However, this equation is
not always appropriate for finite-size droplets, 
especially for droplet of a rather smaller 
size \cite{malila2018monolayer}. Instead, we can compute
the surface tension of a droplet in the following way.
We consider an aqueous droplet with a radius, $r$, 
which can be divided into a surface monolayer of thickness,
$\delta$, and an interior (bulk) with radius, 
$r - \delta$. The droplet surface tension, $\sigma$,
depends on the value of the bulk composition 
$\mathbf{x}^b = (x_1^b, x_2^b, \ldots)$ and the surface
composition $\mathbf{x}^s = (x_1^s, x_2^s, \ldots)$ \cite{malila2018monolayer},
namely:
\begin{equation}
\sigma(\mathbf{x}^b, T) = \frac{\sum_i \sigma_i v_i x_i^s}{\sum_i v_i x_i^s},
\label{EoS}
\end{equation}
where $x_i^b = n_i^b / \sum_j n_j^b$ is the bulk mole
fraction for each species, $i$, in the droplet, $x_i^s$ is
the surface mole fraction, defined in the same manner, and
$v_i$ and $\sigma_i$ are the molecular volumes and surface
tensions of each pure component, $i$, respectively. 
$n_i^s, n_i^b$, and $n_i^t = n_i^s + n_i^b$, are the number
of particles for the $i$th species, for the surface, 
the bulk, and the whole droplet, respectively. 
The total amount of molecules in the monolayer, $n_i^s$, 
are evaluated from the volume of the monolayer, 
$V^s = 4 \pi [r^3 - (r - \delta)^3]/3$, where the thickness,
$\delta$, of the monolayer is given from the 
equation \cite{malila2018monolayer}:
\begin{equation}
\delta = \left(\frac{6}{\pi} \sum_i v_i x_i^s \right)^{1/3}.
\label{delta}
\end{equation}

In particular, considering the three mole fractions for 
water (subscript `1'), salt (subscript `2'), and 
surfactant (subscript `3'), Eq.~(\ref{EoS}) reduces to:
\begin{equation}
\sigma(x_1^b, x_2^b, x_3^b) = \frac{\sigma_1 v_1 x_1^s + \sigma_2 v_2 x_2^s + \sigma_3 v_3 x_3^s}{v_1 x_1^s + v_2 x_2^s + v_3 x_3^s}.
\label{sigma}
\end{equation}
Here, we assume that all surfactant is on the surface of
the droplet until it covers the whole surface and starts
to form aggregates. Moreover, the ratio between salt and
water is the same for bulk, as well as for the surface.
Therefore, the partitioning formulas for the monolayer are: 
\begin{equation}
n_1^s = \frac{V^s - n_3^s v_3}{v_1 + n_2^t v_2 / n_1^t}, \ n_2^s = n_1^s \frac{n_2^t}{n_1^t}, \ n_3^s = \min(n_3^t, V^s / v_3), 
\end{equation}
where 
\begin{equation}
n_1^t = \frac{4\pi}{3} \cdot \frac{r^3-r_d^3}{v_1}, \ n_2^t = (1 - c_{32}) \cdot \frac{4\pi}{3} \cdot \frac{r_d^3}{v_2}, \ n_3^t = c_{32} \cdot \frac{4\pi}{3} \cdot \frac{r_d^3}{v_3},
\label{total}
\end{equation}
$c_{32} = n_3^t v_3 / (n_2^t v_2 + n_3^t v_3)$ is the volume
fraction of surfactant in the atmospheric CCN particles,
and $r_d$ is the dry radius of a super-droplet. 
For simplicity, in our calculations we assumed 
that $r_d \ll r_w$, where $r_w$ is the wet radius of a
super-droplet. Finally, we used a mass fraction, $c$, 
of surfactant, with an appropriate conversion formula:
\begin{equation}
c_{32} = \frac{c m_2 v_2}{c m_2 v_3 - (c-1) m_3 v_2},
\end{equation} 
where $m_2$ and $m_3$ are molecular masses of salt and
surfactant, respectively. 

The model with mass concentrations $c = 0\%$ is the reference
model, that is, 100\% of the dry part is ammonium sulfate.
A simulation with, \textit{e.g.}, $c = 20\%$ stands for 
20\% of SDS and 80\% of ammonium sulfate.
In Ref.\ \cite{facchini1999cloud}, the concentration of
surfactant in the atmospheric CCN is about 5\%. 
However, the proportions of particular types of aerosols
over the globe may vary \cite{facchini1999cloud, facchini2000surface, jimenez2009evolution, prisle2010surfactants, ovadnevaite2017surface}.

Overall the extent of accuracy of some of the model
assumptions for cloud droplets made above are subject 
to current discussion in the field, but the scheme used
constitutes a baseline from which surfactant concentration
effects can be incorporated into a previously
surfactant-free model. We have used the relative surfactant
concentration, $c$, in the atmospheric water
(or a single droplet), as the wet radius of the droplet
evolves. In this case, surfactant concentration in
droplet's water also evolves and does not remain constant
during the simulation \cite{facchini1999cloud}. 
For example, during a droplet growth, the amount of
surfactant may initially exceed the critical aggregation
concentration (CAC), and then, due to the accumulation of
water by the droplet, the concentration may decrease below
the CAC level, with the actual amount of surfactant in 
the droplet remaining the same. However, the amount of
salt and surfactant can change due to droplet collisions.


Equation~(\ref{koehler}) is used to calculate the saturation
ratio by using $\sigma$ as obtained from Eq.~(\ref{sigma}).
In our study, $\sigma$ was iteratively computed by using
Steffensen's method \cite{dahlquist2003numerical} and 
Eqs.\ (\ref{delta})--(\ref{total}). We used wet, $r$,
and dry, $r_d$, radii of super-droplets calculated by the
UWLCM library. In this way, the effect of the surface
tension, as a result of surfactant's presence in the
droplets, is taken into account in the numerical model.
In turn, surface tension shall affect the supersaturation
curve, which eventually impacts the droplet formation 
and evolution, as will be further elaborated below.


\section{Results}
\label{res}

\begin{figure*}
\centering
\includegraphics[width=0.8\textwidth]{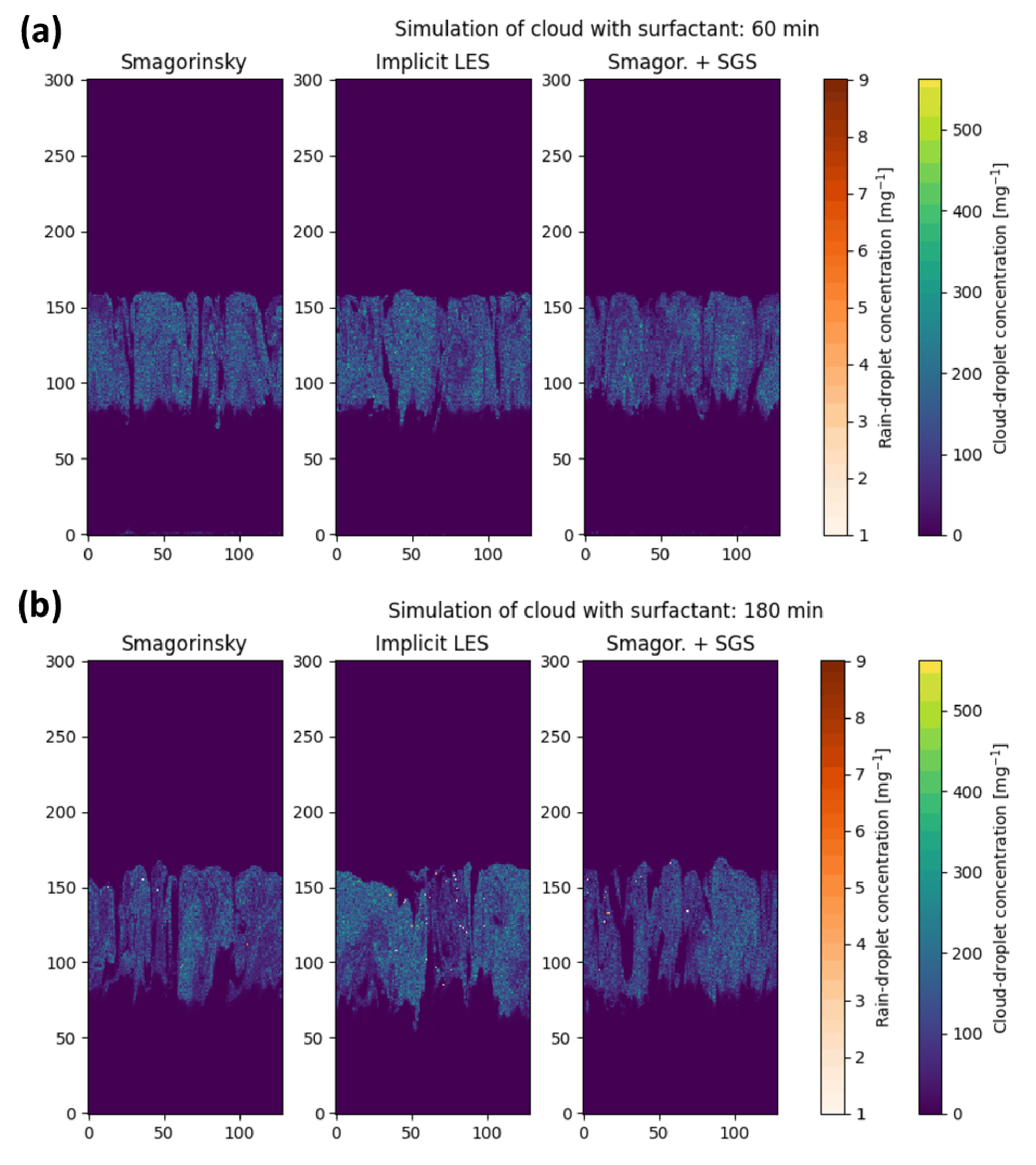}
\caption{Cloud-droplet (0.5 $\mu$m $< r < $ 25 $\mu$m,
purple–yellow scale) and rain-droplet ($r > $ 25 $\mu$m,
white–red scale) concentration (number of droplets per
miligram of dry air) for the snapshots of the SDM simulation
with 25\% of surfactant (mass of surfactant in the droplet
divided by the collective mass of surfactant and salt), 
for three different modeling approaches and times
(a) $t$ = 60 min, (b) $t$ = 180 min, as indicated. 
Horizontal and vertical axes values for each subplot are
counts of grid cells. }
\label{anim}
\end{figure*}

To test the size of the effect produced by the surfactant,
we have reproduced drizzling marine 
stratocumulus \footnote{As stratocumulus is a dominant cloud formation in Earth's atmosphere, our results are considerably general.} 
simulations using the second Dynamics and Chemistry of 
Marine Stratocumulus (DYCOMS-II) second research flight
(RF02) setup as in Ref.\ \cite{dziekan2019university}, 
but with the influence of various surfactant concentrations
added to the reference model. For each examined concentration
of surfactant, we conducted one hundred 6-hour 2D simulations
to average over multiple cloud evolutions. 
We used a grid of a size 6.4~km horizontally and 1.5~km
vertically, with grid cells $50 \times 5$~m each, 
with periodic boundary condition on the left/right boundary
of the grid. The time step for our computations was 1 s (see Ref.\ \cite{dziekan2019university} for more details about choosing appropriate time step value).

Figure \ref{anim} shows the snapshots from the example
simulation with a concentration of cloud and rain droplets.
In general, the Implicit LES approach produces more rain
than the other two approaches,
in line with previous findings \cite{sato2017grid}.
Moreover, to reliably calculate average values for
properties, such as the rain-droplet concentration, 
which are nonlinear and sensitive to initial conditions, 
the realization of a large ensemble of simulations to deal
with possible statistical errors for the derived quantities
were required. We have found that an ensemble of 
100 simulations for each system was enough to yield 
reliable results for the properties of each system.

\begin{figure}
\centering
\includegraphics[width=\columnwidth]{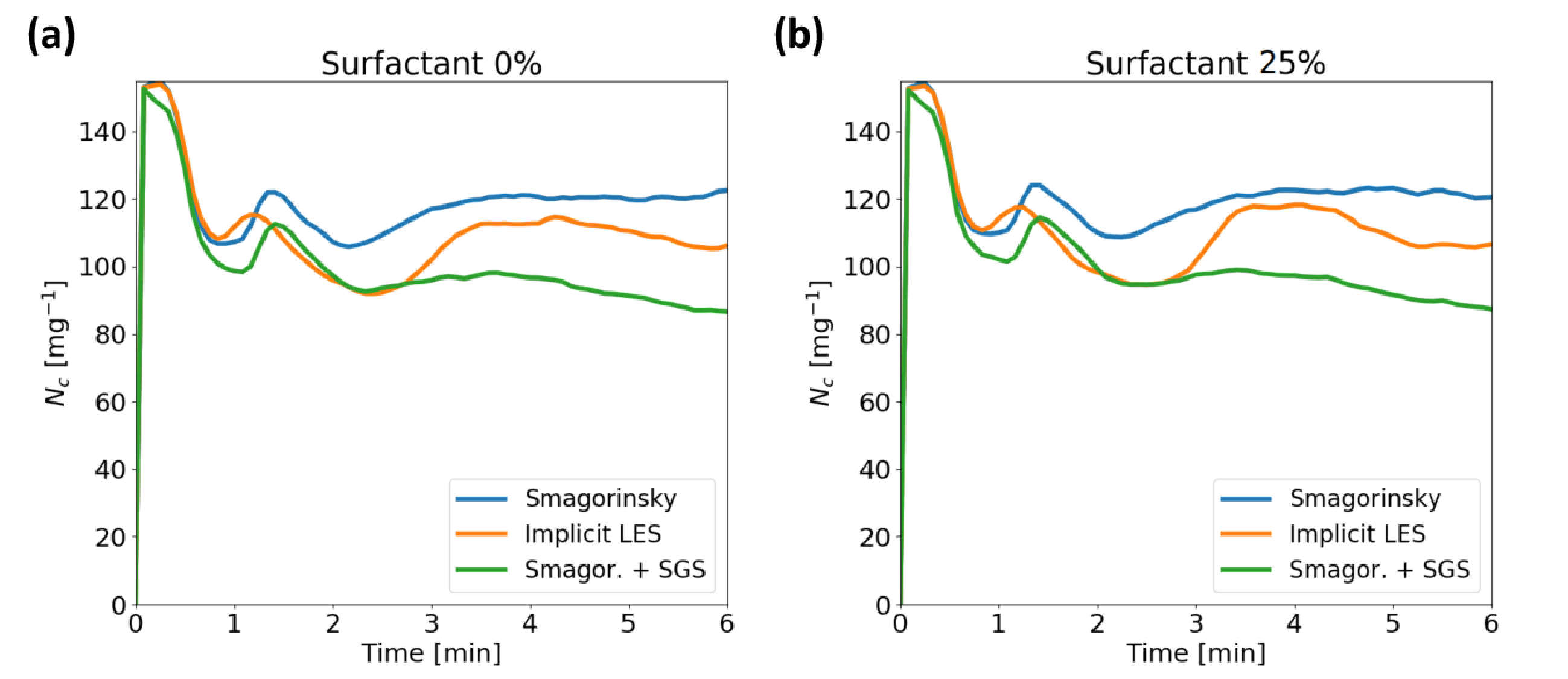}
\caption{Time evolution of concentration of cloud droplets
in cloudy grid cells, $N_c$, averaged for 100 simulations:
original model from Ref.\ \cite{dziekan2019university}
reproduced (a) and our model with 25\% mass ratio
of surfactant (b).}
\label{evol}
\end{figure}

\begin{figure}
\centering
\includegraphics[width=\columnwidth]{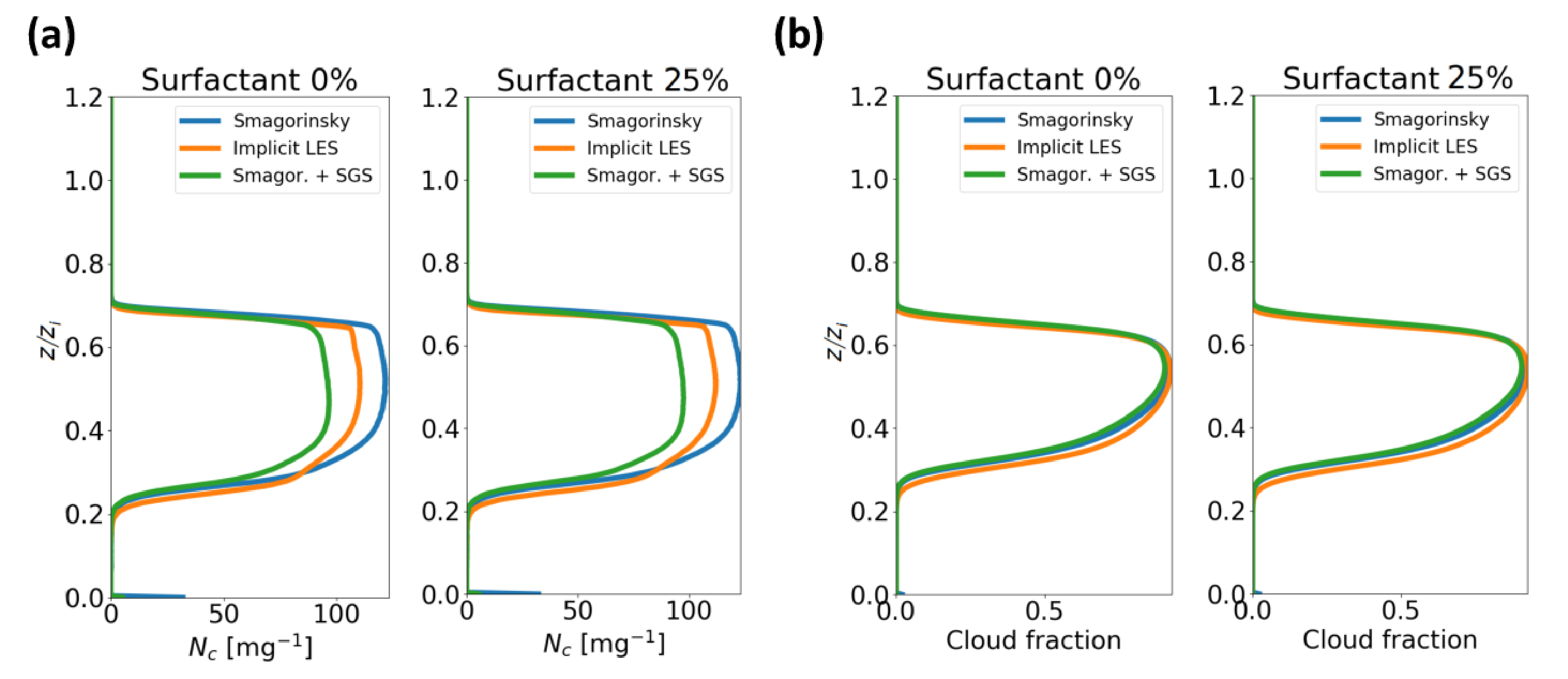}
\caption{Vertical profiles of: (a) concentration of cloud
droplets in cloudy grid cells for the original model
without surfactant \cite{dziekan2019university};
concentration of cloud droplets in cloudy grid cells for our
model with 25\% of surfactant, as indicated; (b) cloud fraction
for the original model; cloud fraction for
our model with 25\% of surfactant, as indicated. 
The plotted results are averages of 100 simulations for
the whole simulation grid, 
the scaled height $z/z_i$ is defined as in Ref.\ \cite{dziekan2019university}.} 
\label{prof}
\end{figure}

Example time evolutions and profiles obtained for the 
model with and without surfactant are shown in 
Figs.\ \ref{evol} and \ref{prof} (cf.\ similar results
in Ref.~\cite{dziekan2019university}). Cloud droplets are
liquid particles with a radius in the range of
0.5 $\mu$m < $r$ < 25 $\mu$m. Cloudy cells are defined as
those with a concentration of cloud droplets greater than
20 cm$^\text{-3}$. The cloud fraction is the ratio of cloudy
cells to the total number of cells. The concentration of
cloud droplets in cloud grid-cells (Fig.~\ref{evol})
initially grows rapidly, then drops down due to the
evaporation process, and finally exhibits slight
fluctuations. In the case of the `Smagorinsky + SGS' scheme
(SGS motion of super-droplets) we observe a decrease in the
concentration of cloud droplets as the simulation progresses,
whereas in the case of the Smagorinsky scheme (SGS only for
Eulerian variables), there is a slight increase with some
fluctuations. The profiles of Fig.~\ref{prof} indicate the
greatest concentration of cloud droplets in the case of the
Smagorinsky scheme, in contrast to the lowest values observed
in the case of the `Smagorinsky + SGS' scheme. However,
we see almost no visible differences between the different
models both in the case of systems with and without
surfactant. As for the cloud cover, we have obtained the 
same result for all schemes, also, independently of the
presence of surfactant in the systems. Our results indicate
that the effect of surfactant for these properties is
negligible, as evidenced by our results in
Figs.~\ref{evol} and \ref{prof}.

\begin{figure}
\centering
\includegraphics[width=\columnwidth]{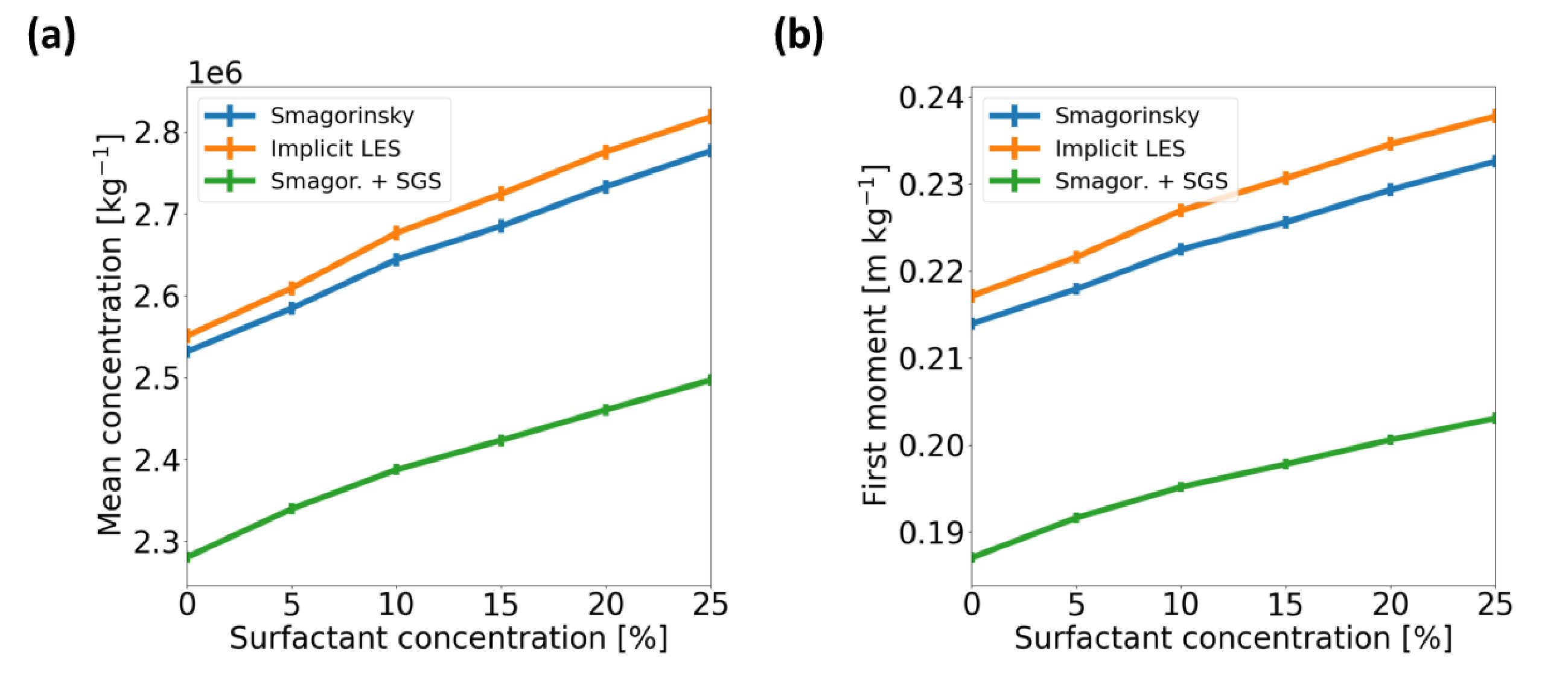}
\caption{Number of activated droplets
(critical supersaturation exceeded) per kilogram of dry
air \textit{vs.} surfactant concentration $c$ (a), and
1st moment of the dry spectrum [defined by Eq.\ (29)
in Ref.\ \cite{arabas2015libcloudph++}, b] \textit{vs.}
surfactant concentration $c$, averaged over 100 six-hour
simulations, for three different modeling approaches,
as indicated. Error bars (vertical) show the statistical errors of the calculated mean quantities for each surfactant concentration. }
\label{actRH}
\end{figure}

There are, however, properties that are significantly
affected by surfactant in our model. 
These properties are the number of activated
particles\footnote{Before the activation the growth of the
particles is stable with a particular equilibrium radius,
after the activation it is unstable, and cloud droplets
are formed.} (referring to those with the 
critical supersaturation exceeded), as well as their size.
Both these values increase with an increasing surfactant
concentration, independently of the modeling approach
(Fig.~\ref{actRH}).
Although including the SGS motion of the Lagrangian particles
(super-droplets) will lead to a decrease of the concentration
and size of the activated droplets, we can observe that the
same positive dependence on surfactant concentration still
remains. Moreover, there is no saturation of this effect
with a growing surfactant concentration. Finally, we 
have observed a slight decrease in the speed of the
concentration and the pace of growth for all modeling
approaches.  

\begin{figure}
\centering
\includegraphics[width=\columnwidth]{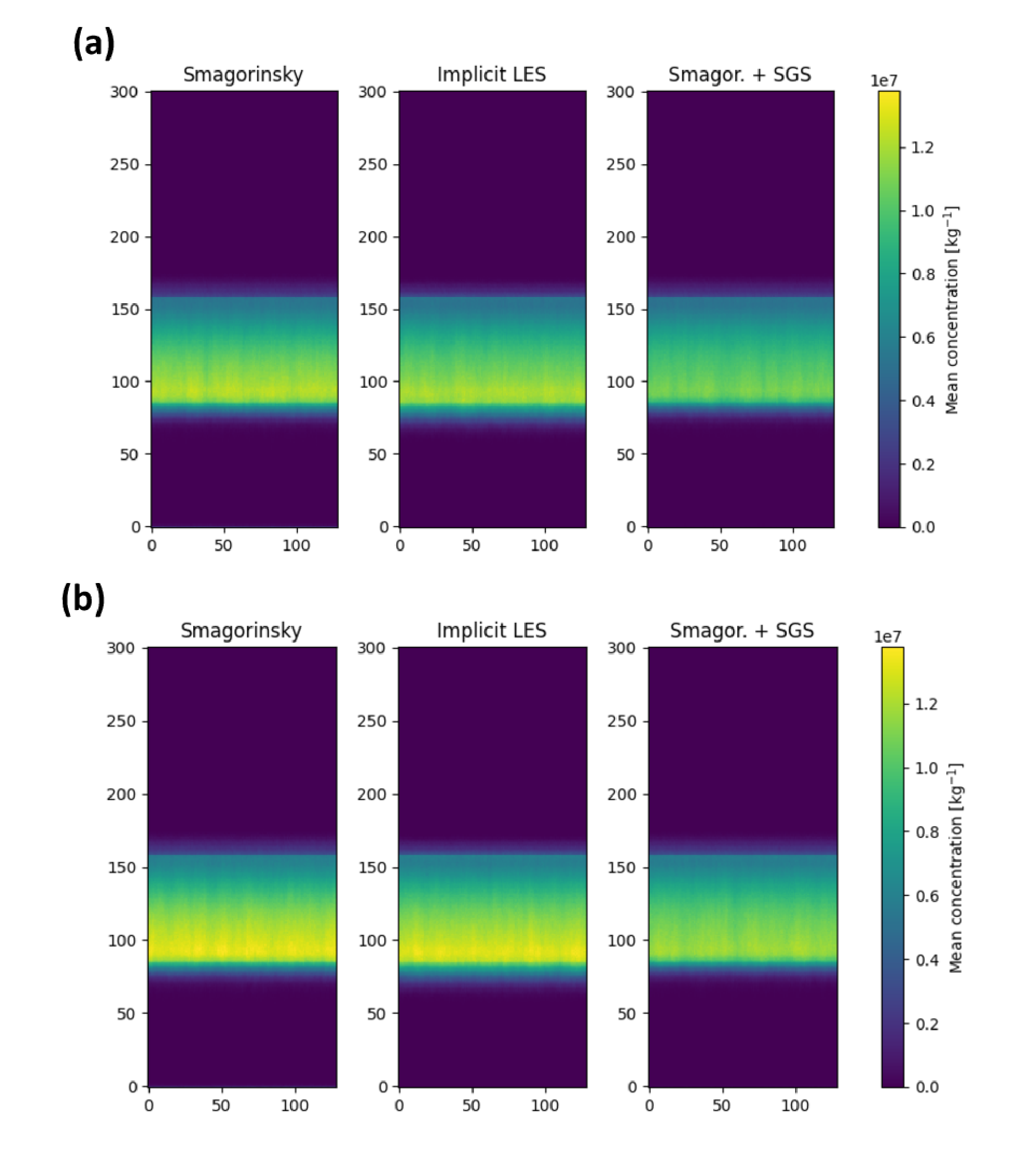}
\caption{Maps of the number of activated droplets per
kilogram of dry air for the whole simulation grid, 
averaged over 100 simulations for the same sets of the
model parameters and over six-hour time of each simulation,
for three different modeling approaches, as indicated. 
(a) reference model (without surfactant).
(b) model with 25\% of surfactant. Note that the
maps are expanded by ten times in a vertical direction, 
for a better view.}
\label{actmap}
\end{figure}

Another key observation from our calculations concerns the
biggest difference between the number of activated droplets
for the reference model and the model with high surfactant
concentration, which occurs at the bottom of a
cloud (Fig.~\ref{actmap}). 
In contrast, the largest droplets are concentrated at the
top of the cloud (Fig.~\ref{actmap}). Hence, our results
indicate that ``new'' droplets are activated at the bottom,
and then they grow moving up the cloud. In a sense, 
there is a circulation of droplet through an activation and
growth process from the bottom of the cloud towards the top.

\begin{figure}
\centering
\includegraphics[width=0.5\columnwidth]{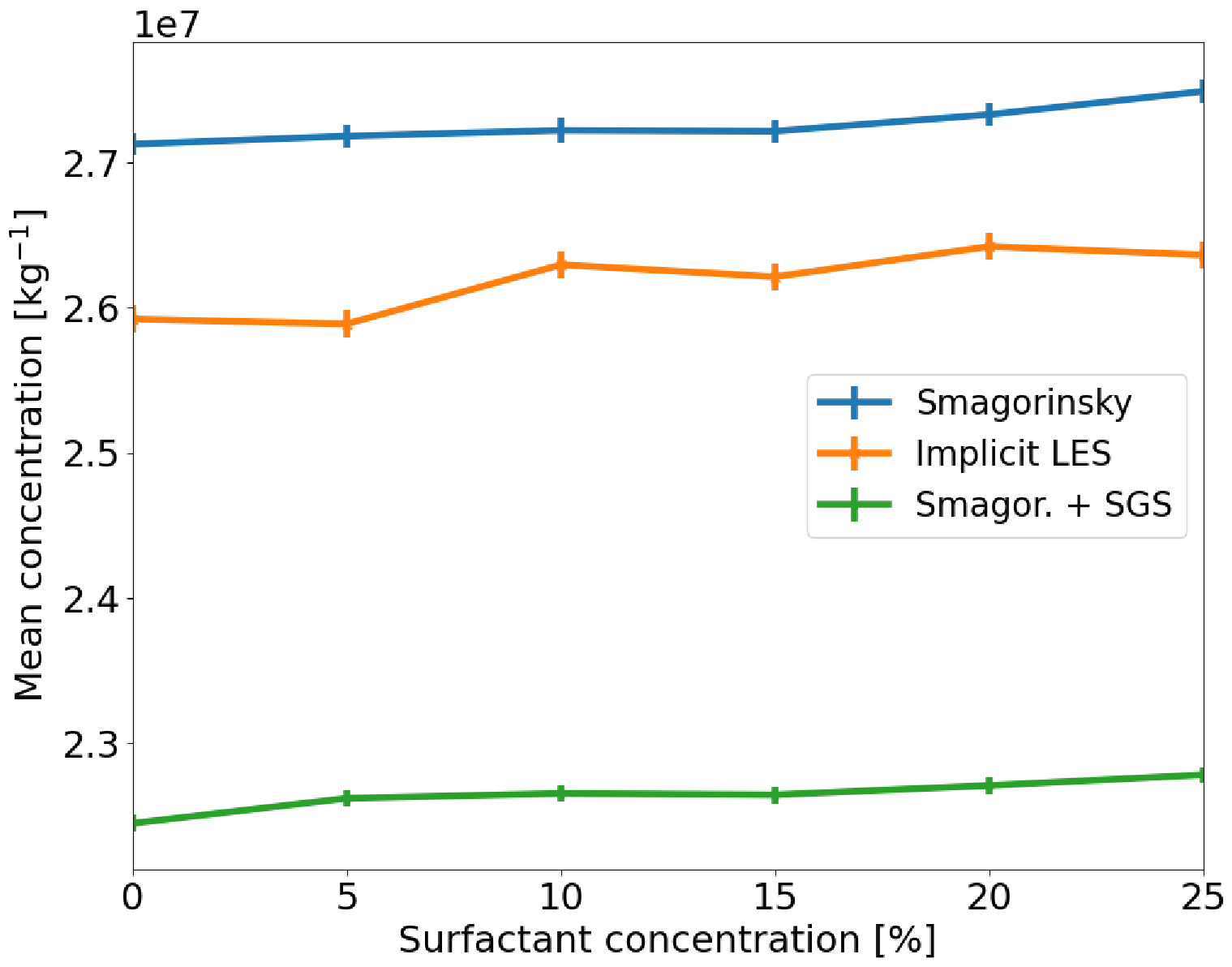}
\caption{Concentration of droplets that exceed the critical
radius for the K\"ohler curve (``radius activated'') per
kilogram of dry air \textit{vs.} surfactant concentration
$c$, averaged over 100 six-hour simulations, for three
different modeling approaches, as indicated. Error bars (vertical) show the statistical errors of the calculated mean quantities for each surfactant concentration. }
\label{actrw}
\end{figure}

If ``radius activated'' droplets were to be identified by
checking whether their radius exceeds the critical value for the K\"ohler curve [Eq.~(\ref{koehler})] instead
of using the critical supersaturation as the threshold, 
the increase with a growing surfactant concentration becomes
smaller (Fig.~\ref{actrw}). 
This is due to the fact that the critical radius is also
exceeded by droplets that were activated a long time ago,
while critical supersaturation is exceeded usually only once,
when the actual activation happens. Therefore, the absolute
concentration is larger when the critical radius is exceeded,
because it also takes into account the ``old '' cloud
droplets, while when the critical supersaturation is
exceeded, only the ``new'' cloud droplets are considered.
Hence, the relative increase with growing surfactant
concentration is bigger in the latter case. We have also
checked that the absolute differences in the number of
droplets for these two cases are of the same order of
magnitude, which further confirms our conclusions.
However, the Smagorinsky scheme produces
more `radius activated'' droplets than the ILES scheme, 
which is opposite to the case of ``supersaturation''
activated droplets in Fig.~\ref{actRH}; 
the `Smagorinsky + SGS' scheme still produces the smallest
amount of droplets for both measures.  

\begin{figure}
\centering
\includegraphics[width=\columnwidth]{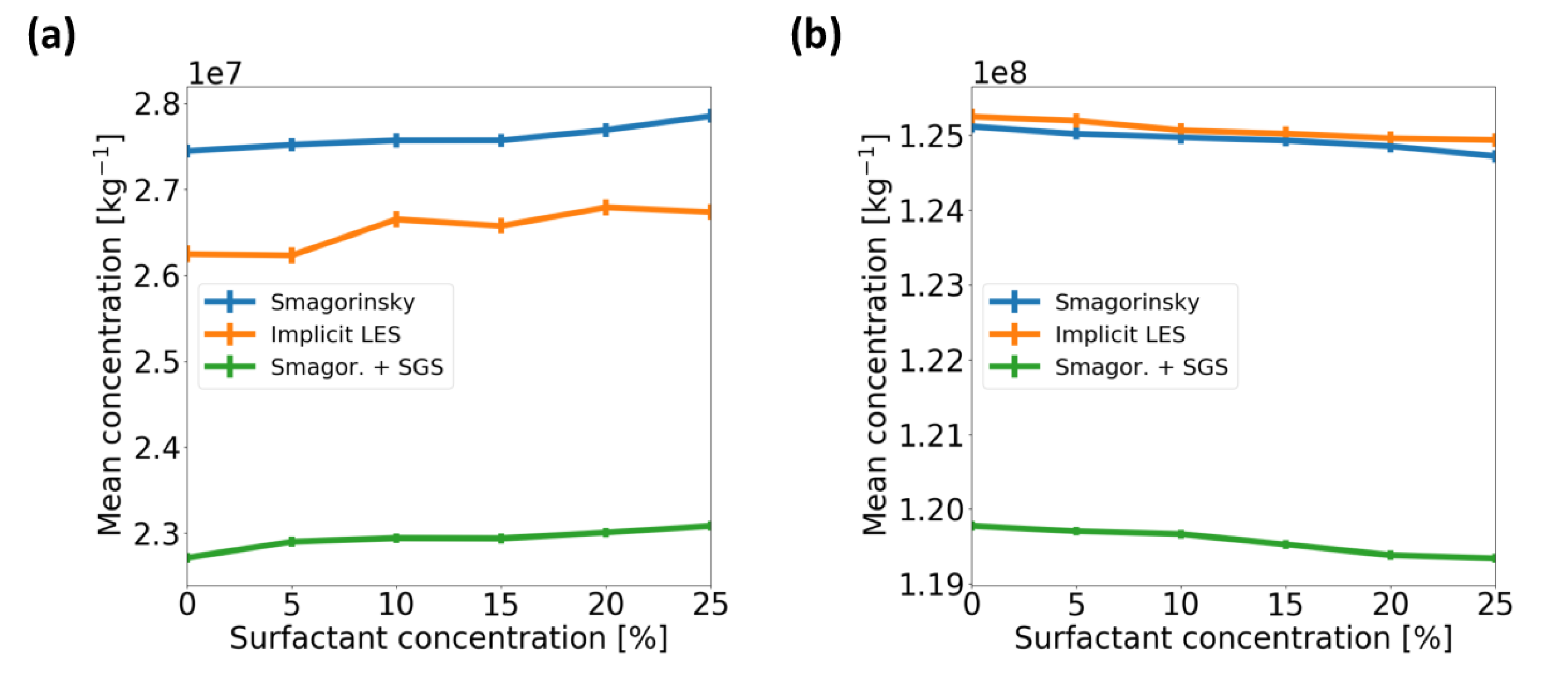}
\caption{Concentration of cloud droplets 
($0.5$ $\mu$m $< r < $ 25 $\mu$m, a) and aerosol 
particles ($r < 0.5$ $\mu$m, b) \textit{vs.} 
surfactant concentration $c$, averaged over 100 six-hour
simulations, for three different modeling approaches,
as indicated. Error bars (vertical) show the statistical errors of the calculated mean quantities for each surfactant concentration. }
\label{cloud}
\end{figure}

The concentration of cloud droplets remains almost 
identical as in the case of the droplets with radius
exceeding the critical radius, because activated droplets
are converted into cloud droplets (Fig.~\ref{cloud}).
There is also a decrease of aerosol concentration as
surfactant concentration increases (Fig.~\ref{cloud}, 
lower panel).
Such an effect can be explained by the fact that aerosol
particles transform into cloud droplets, which leads to an
anti-correlation between the latter two effects. 
However, the differences between the Smagorinsky and the
ILES schemes are negligible for the concentration of
aerosol particles, while for the concentration of cloud
particles the Smagorinsky scheme yields more particles.

\begin{figure}
\centering
\includegraphics[width=\columnwidth]{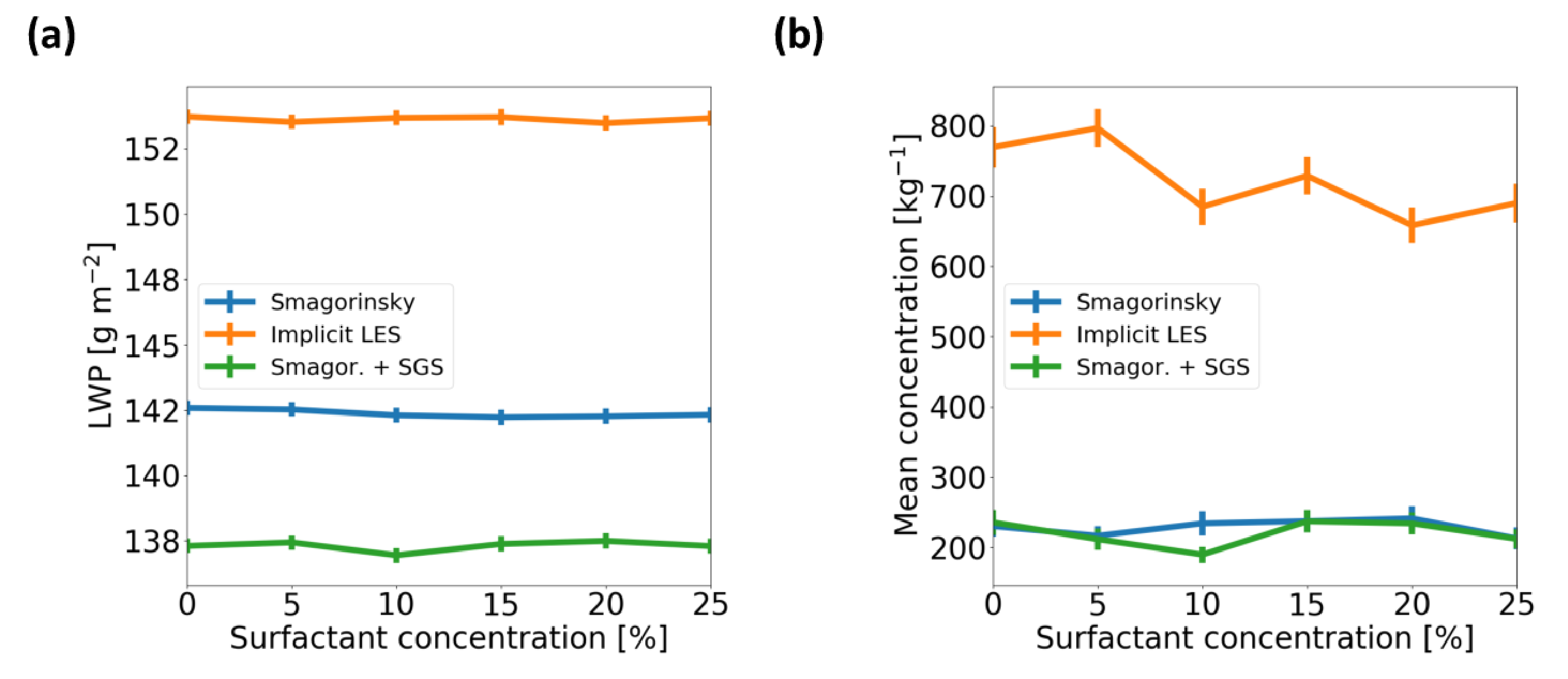}
\caption{Liquid water path (a) and concentration of rain
drops ($r > 25$ $\mu$m, b) vs. surfactant concentration
$c$, averaged over 100 six-hour simulations, for three
different modeling approaches, as indicated. Error bars (vertical)
show the statistical errors of the calculated mean quantities for
each surfactant concentration. }
\label{LWP}
\end{figure}

Finally, we have found that the amount of water in the 
cloud and the concentration of rain drops do not change 
with the concentration of surfactant in the case of all
different models (Fig.\ \ref{LWP}). 
Still, the Implicit LES approach produces more water and
rain droplets in the cloud than the other two approaches.
Moreover, there is a big discrepancy between the
precipitation results for different approaches. 
Such differences have also been previously reported in
Ref.~\cite{dziekan2019university}. In the case of ILES,  
a decrease in the concentration of rain droplets with a
growing surfactant concentration is observed.
However, in view of the large statistical errors, 
this effect deserves further investigation. 

The quantities shown in Figures~\ref{actrw}-\ref{LWP} do not appear to show as strong a change with concentration as Fig.~\ref{actRH}, but this is partly due to an optical effect caused by the large differences in values predicted by the three particular models that we used (Smagorinsky, Implicit LES, and Smagor. + SGS) which are shown together on the same plots. The effects of the growing surfactant concentration relative to 0\% concentration show the same dependence for all three models that we have used. We did not plot other quantities that showed no clearly discernible surfactant effect. The uncertainty in the data points are primarily statistical. It is caused by the large shot-to-shot fluctuations in the quantities between particular simulation runs that have different noise histories. The data points shown are calculated after averaging the results over different noise realizations. Other sources of errors are smaller and can be neglected.

\section{Conclusions and Discussion}
\label{concl}


In this study, we have proposed a model for a cloud
consisting of surfactant-laden droplets, which is based on
the Lagrangian particle-based microphysics scheme
super-droplet method \cite{shima2009super}. The model is
based on the modification of the Kelvin term
(the term responsible for the droplet surface behavior) 
in K\"ohler's equation (\ref{koehler}) by introducing
Eq.\ (\ref{sigma}) to calculate surface tension, $\sigma$,
with the presence of surfactant in the droplet.
We have shown that atmospheric surfactants influence the
cloud formation process by increasing the concentration 
and size of activated droplets. This result is in accordance
with previous studies suggesting that a decrease in surface
tension of cloud droplets due to surfactant might lead to
an increase in the population of the smaller-size 
droplets \cite{latif2004surfactants}, especially in the
bottom of the clouds. Although the decrease of surface 
tension due to surfactant is limited by CAC, we have not
observed a saturation of the effect of the droplet
concentration increase. However, it might be the case that
such saturation occurs for higher surfactant amounts than
what has been assumed here, which calls for further studies.
We have also found no clear link between the presence of
surfactant in the system and the amount of rain. 
This is also a point that might require further investigations
in the future. In this direction, further developments of the
current model are essential to obtain a more comprehensive
picture of the underlying phenomena. Possible extensions
might include: (a) improving surfactant partitioning or the
interpolation between partial surface tensions in (\ref{EoS})
(these details remain an active topic of investigation 
in the field);
(b) adding the influence of surfactant to the Raoult term 
(for example, see Ref.\ \cite{petters2013single});
(c) introducing coalescence efficiency that changes in the
presence of surfactant \cite{hudson2003effect, leal2004flow};
(d) other types of surfactants than SDS;
(e) presence of ice particles and other types of 
clouds \cite{grabowski2019modeling, shima2020predicting}; (f) examination of other grid sizes and time steps, with a possibility to apply a \emph{substepping} procedure \cite{dziekan2019university}; (h) study other properties of clouds than those shown in the paper.
Such developments might lead to a more comprehensive model of
a cloud in the presence of surfactant, which will enable the
modeling of various aspects of Earth's climate, for instance,
regional changes in precipitation, or the exact role of
aerosols in cloud formation.

Our numerical model offers new opportunities for 
investigating the role of
surfactants at macroscopic models that involve topological
changes of droplets (\textit{e.g.} coalescence of 
surfactant-laden droplets). In this context, such models
can be further refined based on information from 
molecular-scale models \cite{Theodorakis2015a,Smith2018,theodorakis2015modelling,Theodorakis2019,theodorakis2019molecular}. Hence, the
SDM holds promise for further developments in the area
of numerical modeling for large-scale systems consisting
of droplets with different properties, as has been shown
here in the case of a cloud with surfactant-laden droplets.
We anticipate that our study will motivate further work in
this direction.

\begin{acknowledgments}
This research has been supported by the National Science Centre, Poland, under grant No.\ 2019/34/E/ST3/00232. This research was supported in part by PLGrid Infrastructure. MD is grateful for inspiring discussion and valuable comments made by Jussi Malila and Piotr Dziekan. 
\end{acknowledgments}

\section*{Data Availability Statement}
The data that support the findings of this study are available from the corresponding author upon reasonable request.

\providecommand{\noopsort}[1]{}\providecommand{\singleletter}[1]{#1}%

\end{document}